\documentclass[conference]{IEEEtran}
\usepackage[dvipsnames]{xcolor}
\makeatletter

\def\ps@IEEEtitlepagestyle{%
  \def\@oddfoot{\mycopyrightnotice}%
  \def\@evenfoot{}%
}
\def\mycopyrightnotice{%
  {\footnotesize 979-8-3315-3559-9/25/\$31.00~\copyright~2025 IEEE\hfill}
  \gdef\mycopyrightnotice{}
}

\usepackage{blindtext}
\usepackage{eso-pic}
\IEEEoverridecommandlockouts
\usepackage{cite}
\usepackage{amsmath,amssymb,amsfonts}
\usepackage{algorithmic}
\usepackage{graphicx}
\usepackage{textcomp}
\usepackage{xcolor}
\def\BibTeX{{\rm B\kern-.05em{\sc i\kern-.025em b}\kern-.08em
    T\kern-.1667em\lower.7ex\hbox{E}\kern-.125emX}}
    
\usepackage{eso-pic}
\newcommand\AtPageUpperMyright[1]{\AtPageUpperLeft{%
 \put(\LenToUnit{0.17\paperwidth},\LenToUnit{-2cm}){%
     \parbox{0.9\textwidth}{\raggedleft\fontsize{8}{11}\selectfont #1}}%
 }}%
\newcommand{\conf}[1]{%
\AddToShipoutPictureBG*{%
\AtPageUpperMyright{#1}
}
}    

\usepackage{graphicx}
\usepackage{textcomp}
\usepackage[dvipsnames]{xcolor}
\usepackage{subcaption}
\usepackage{makecell}
\usepackage{diagbox}
\usepackage{arydshln}
\usepackage{xspace}

\usepackage{etoolbox}

\AfterEndEnvironment{figure}{\vskip-0.4ex}

\begin{document}

\title{\vspace*{0.75cm} Energy-Aware LLMs: A step towards sustainable AI for downstream applications\\
\thanks{This work was supported by NSERC (under project ALLRP 566589-21) and InnovÉÉ (INNOV-R) through the partnership with Ericsson and ECCC.}
}

\author{\IEEEauthorblockN{1\textsuperscript{st} Nguyen Phuc Tran}
\IEEEauthorblockA{\textit{Computer Science and Software Engineering} \\
\textit{Concordia University}\\
Montréal, Québec, Canada \\
phuc.tran@mail.concordia.ca}
\and
\IEEEauthorblockN{2\textsuperscript{nd} Brigitte Jaumard}
\IEEEauthorblockA{\textit{Computer Science and Software Engineering} \\
\textit{Concordia University}\\
Montréal, Québec, Canada \\
brigitte.jaumard@concordia.ca}
\and
\IEEEauthorblockN{3\textsuperscript{rd} Oscar Delgado}
\IEEEauthorblockA{\textit{Systems Engineering} \\
\textit{École de Technologie Supérieure}\\
Montréal, Québec, Canada \\
oscar.delgado@etsmtl.ca}
}

\maketitle
\conf{\textit{  V. International Conference on Electrical, Computer and Energy Technologies (ICECET 2025) \\ 
3-6 July 2025, Paris-France}}
\begin{abstract}
Advanced Large Language Models (LLMs) have revolutionized various fields, including communication networks, sparking an innovation wave that has led to new applications and services, and significantly enhanced solution schemes.
Despite all these impressive developments, most LLMs typically require huge computational resources, resulting in terribly high energy consumption.

Thus, this research study proposes an end-to-end pipeline that investigates the trade-off between energy efficiency and model performance for an LLM during fault ticket analysis in communication networks.
It further evaluates the pipeline performance using two real-world datasets for the tasks of root cause analysis and response feedback in a communication network.

Our results show that an appropriate combination of quantization and pruning techniques is able to reduce energy consumption while significantly improving model performance.
\end{abstract}



\section{Introduction}

Given the striking marks LLMs attain in natural language processing (NLP), mining textual, and generation tasks, they have been a new emerging possibility of applying AI in several domains.
In fact, it is developing with many complexities owing to high performance; hence, it develops substantial advances within several networking areas in communication networks \cite{soman2023observations}.
Their downstream applications could be root cause analysis (RCA) \cite{chen2024automatic}, question and answer system \cite{roychowdhury2024unlocking}, etc.
On the downside, this also presents a costly drawback that requires significant computing resources \cite{patel2024characterizing}.
In addition, the growth of LLMs in size of parameters and complexity is accompanied by a parallel increase in energy consumption, posing worries about AI's energy footprint and ecological footprint.
On top of that, AI is getting bigger and better, but it is also becoming more expensive to run, especially because of the energy it needs, especially for big language models.
Therefore, the huge amount of energy needed to build and operate these models is a huge challenge for computing resources.

Today, the development and operation of LLMs are deeply related to the capabilities of the underlying communication networks \cite{maatouk2024large}.
These models require massive datasets, not limited to one location but distributed across the local network and from external networks, requiring robust and high-bandwidth networking.
For instance, when training an LLM often involves processing a huge of text data, which demands substantial network bandwidth.
Besides this, during the time of inference, the system is highly dependent on effective data transfer from the model to the users.
Thus, the energy consumption associated with communication networks is a significant factor in the overall environmental impact of AI systems.
It includes data transmission, routing, and processing within these networks, which consume substantial amounts of electricity.
In fact, these LLMs are often deployed in data centers which are the backbone of modern communication networks and are generally well known for their very high energy consumption due to complex computing requirements, operating resources, and network equipment.
Thus, LLM optimization is fundamental to reducing the overall footprint of energy consumption in these data centers and communication networks and hence contributes to sustainable AI.

Indeed, finding a compromise between boosting LLM performance and minimizing energy consumption is now in addition a top priority in the system operation.
In this paper, we investigate and address the compromise between the energy footprint and performance of LLMs in downstream applications deployed in a data centre environment.
In detail, our contributions can be summarized as follows:
\begin{itemize}
    \item Propose an end-to-end pipeline to evaluate the compromise of LLM energy efficiency and model performance using quantization and pruning techniques.

    \item Evaluate our pipeline on the two real-world datasets in the communication network.

    \item Provide the valuation insights to build sustainable AI applications, particularly using LLM.
\end{itemize}

\section{Related works}

Along with growing interest in LLMs, various techniques have emerged to boost model performance and optimize their efficiency.
Commonly, one technique that helps reduce memory footprint is quantization, which involves lowering the precision of model weights.
Most of these methods \cite{9251854, hu2022lora} aim to represent weights with few representatives.
In the same space, Li \textit{et al.} \cite{10400181} investigated a co-design of quantization and hardware architecture aimed at optimizing quantization performance.
These approaches not only reduce the memory footprint but also preserve model performance.
Although the above works successfully to reduce the memory footprint of LLMs, one key factor is missing in this equation: the energy footprint evaluation. 
While this was important for memory optimization, understanding energy efficiency is equally important, especially once the LLMs find applications in energy-limited settings.
Without considering energy efficiency, improvements in memory alone may not fully reflect the real-world impact of these techniques in terms of sustainability and scalability.

On the other hand, pruning techniques, including both unstructured and structured pruning, are well-known methods for optimizing memory consumption in deep-learning models.
Anwar \textit{et al.} \cite{anwar2017structured} proposed structured pruning for deep convolutional neural networks, where whole filters or channels are removed from the network.
This method results in more efficient models while preserving performance better than unstructured pruning.
Similarly, several works mentioned in the survey \cite{he2023structured} have explored various structured pruning techniques for convolutional neural networks.
Since LLMs are a new type of combination neural network, these pruning techniques have also recently been applied to LLMs \cite{frantar2023sparsegpt}. 
As the same concerns above, these methods have not considered the energy footprint in their experiments, raising a question about the sustainability of AI.

Despite the rapid advancements and widespread adoption of LLMs in various domains, there remains a significant gap in the literature concerning the trade-off between energy efficiency and model performance.
Most existing studies have primarily focused on optimizing memory consumption through techniques such as quantization and pruning.
However, these studies often overlook the critical aspect of energy consumption, which is becoming increasingly important as the environmental impact of AI systems gains attention.
In the context of downstream applications, this gap is even more pronounced, e.g., in communication networks.
Yet, there is a lack of comprehensive studies that evaluate how energy-efficient techniques can be applied to LLMs without compromising their performance in these specific environments.
Our research addresses this gap by proposing an end-to-end pipeline that integrates both quantization and pruning techniques to evaluate and optimize the energy efficiency and performance of LLMs during fine-tuning and inference phases.
By focusing specifically on RCA tasks in communication networks as a use case, we offer a novel perspective that highlights the importance of balancing energy consumption with traditional performance metrics.
This dual focus not only contributes to the advancement of sustainable AI systems but also provides actionable insights for deploying energy-efficient LLMs in practical, real-world scenarios.

\section{Pipeline implementation}

\subsection{Dataset and task}

\begin{figure}[ht]
    \centering
    
    \begin{subfigure}[t]{0.48\columnwidth}
        \includegraphics[width=1\columnwidth]{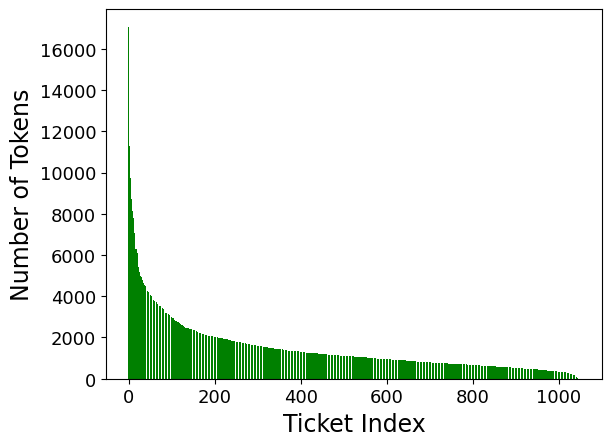}
        \caption{RCA Dataset.}
    \end{subfigure}
    ~
    \begin{subfigure}[t]{0.48\columnwidth}
        \includegraphics[width=1\columnwidth]{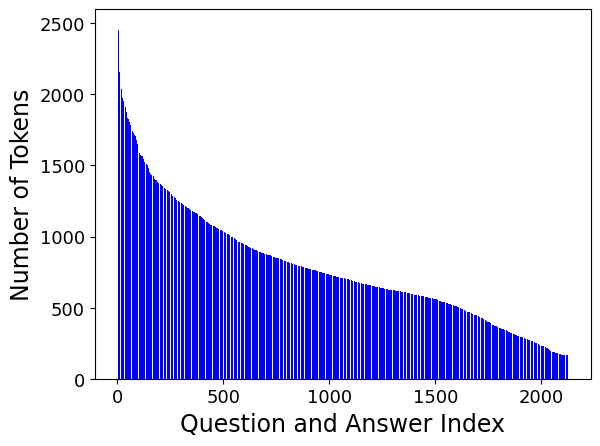}
        \caption{SAF Dataset.}
    \end{subfigure}
    
    \caption{Distribution of token lengths in datasets.}
    \label{fig:dataset-tocken-stats}
\end{figure}

In this research, we utilize two different datasets
for evaluating the trade-offs between the energy footprint and model performance.
Their primary focus is on communication networks to demonstrate a specific downstream application.
For details, they are a private dataset designed for RCA using resolved support tickets and a public dataset, the Short Answer Feedback (SAF) dataset which was published by Filighera \textit{et al.} \cite{filighera-etal-2022-answer}.
These datasets were chosen to provide a diverse evaluation of model performance, each of them focuses on a downstream application (task) within communication networks.
Furthermore, the token length distributions across the datasets, as illustrated in Fig.~\ref{fig:dataset-tocken-stats}, offer critical insights into the model’s ability to handle real-world complexity in downstream tasks.
These variations highlight the differing linguistic and structural challenges posed by each dataset, underscoring the need for adaptive optimization strategies in LLM deployments.

The RCA dataset is used to support root cause analysis in automation networks, a critical task in communication networks and was conducted by 1049 distinct resolved support tickets in 13 network issue categories.
This dataset contains system data derived from network anomaly logs, symptoms, operational events and analysis/solution from technical experts, where each instance represents a specific incident in the network, such as loss signal or performance degradation.
Therefore, the specific task of the LLM is to identify and provide recommendation solutions to the underlying cause of the issue, using input features such as error codes, incident symptoms, network logs, and system metadata.
The other dataset, SAF, combined 2,981 examples including questions, answers, and feedback.
It is designed to address the understanding task of LLM in communication networks, where the goal is to provide answers and the answer feedback on LLM responses to relative questions.
Thus, the task for the LLM in this dataset is to generate relevant answers and feedback that align with the input question.

\subsection{Quantization technique}

In essence, quantization reduces the number of bits to represent each weight, thereby compressing the model.
It should have achieved the goal of faster inference with less use of memory.
However, how quantization, energy efficiency, and model performance relate to a downstream application remains an open area of investigation.
In this work, LoRA \cite{hu2022lora} with BitsAndBytes library
will be our initial chosen approach as a quantization technique candidate to tackle the problems of energy footprint and model performance.
LoRA is the applied technique for pre-trained LLM optimization; therefore, it becomes very productive, reducing memory consumption without sacrificing performance.
Following Xia \textit{et al.} \cite{xia2024quant} findings in their research, our experiment will use $2^n$ bits, where $n$ ranges from $[2, 5]$, as a memory-friendly access for GPUs.
This prioritizes performance and memory efficiency with the ultimate intention of keeping the performance of the LLM as energy-efficient as possible.

\subsection{Pruning techniques}

Pruning is an effective technique for reducing LLM memory footprint \cite{sunsimple} by typically identifying and removing unimportant (or less important) or redundant weights (connections) within the neuron network of an LLM.
As a result, the model can reduce the number of parameters, decrease the model size and reduce computational complexity, leading to faster inference speeds and reduced memory footprint.
Therefore, several pruning methods have been explored including magnitude-based pruning and structured pruning.
Magnitude-based pruning involves removing weights with small absolute values, assuming that these weights have a minimal impact on performance.
On the other hand, structured pruning removes entire units (e.g., filters, neural networks, heads, etc) from the LLM network and often follows a specific pattern.
In practice, the pruning technique can be applied at different stages, such as pre-training, fine-tuning, or even during inference.
Pruning can be a powerful tool, but it is important to use it carefully to avoid sacrificing accuracy.
Excessive pruning can lead to a significant degradation in model performance.
Indeed, pruning techniques can achieve energy benefits, but it is crucial to carefully consider the balance among the pruning ratio, energy efficiency, and model performance.

In our pipeline, these pruning techniques will be applied after being fine-tuned to ensure that our LLM has adapted to the downstream application task.
In addition, this allows LLM to first adjust its parameter weights which align with our dataset, then we apply pruning without negatively impacting its accuracy as much as if we pruned beforehand.
This approach allows us to maximize the LLM performance on our specific task while achieving the benefits of model compression and computational efficiency, making it more suitable for deployment in resource-constrained environments and, therefore, gaining energy efficiency.

\subsection{End-to-end pipeline}

\begin{figure}[ht]
    \centering
    \includegraphics[width=.85\columnwidth]{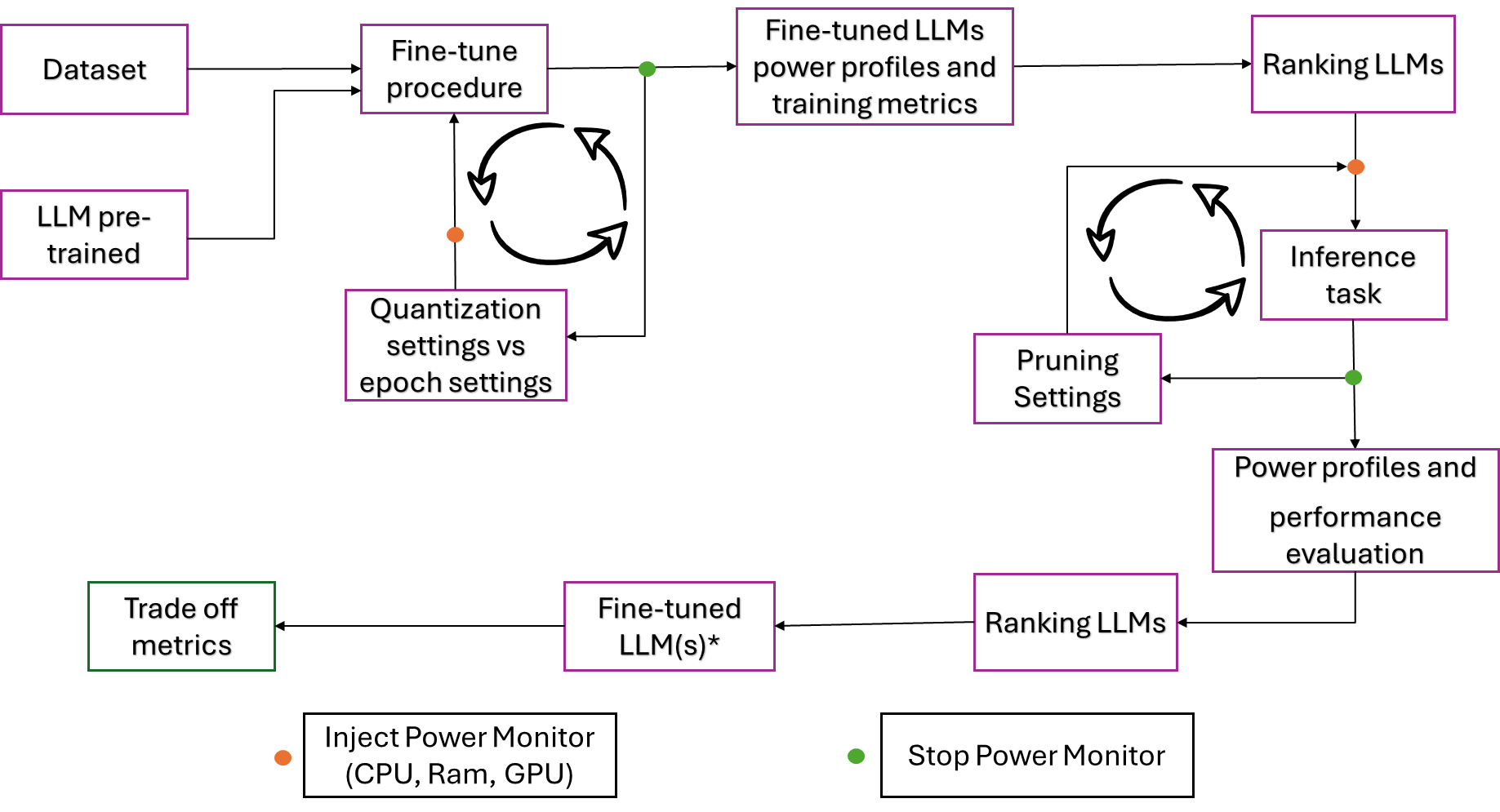}
    \caption{Energy-Performance pipeline evaluation for LLMs.}
    \label{fig:pipeline}
    \vskip-15pt
\end{figure}

Our pipeline, illustrated in Fig.~\ref{fig:pipeline}, integrates the above techniques to address energy footprint and performance challenges in LLMs.
We experiment with different fine-tuning and inference settings to identify the most effective approach.
In this pipeline, a pre-trained LLM will be employed as the baseline LLM for fine-tuning and evaluation.
Throughout the literature review in survey \cite{bannour2021evaluating}, the Code Carbon 
library will be integrated into the pipeline as it is effective in energy measurement for AI/Machine Learning models.

During the fine-tuning phase, we experiment with different quantization settings (e.g., 32 bits, 16 bits) and the number of epochs (e.g., 5 epochs, 10 epochs).
Then, each combination will be evaluated (based on fine-tuning dataset and evaluating dataset), and the power consumption of the CPU, RAM, and GPU will be monitored simultaneously.
These steps help create a comprehensive dataset of metrics, allowing us to analyze the relationship among quantization settings, epochs, and energy consumption.
Thus, we can then identify the balanced combination of these factors to balance energy efficiency and model performance for the fine-tuned phase.
At the end of the first loop, a collection, denoted as $C_{LLM}$, of fine-tuned LLMs and their metrics will be collected and evaluated as inputs for the next loops.
Depending on the pipeline input setting, the top $k$ models that satisfy the weight between model performance and energy efficiency will be selected and transferred to the next phase of the pipeline.
The ranking model based on the weight between model performance, denoted as $\rho$, and energy efficiency, denoted as $\varphi$, for each model in $C_{LLM}$ will be defined as Formula:
\begin{equation}
    R(model_i) = w \cdot \varphi(model_i) + (1-w) \cdot \rho(model_i)
\end{equation}
where: $model_i$: the $i^{th}$ model in the collection $C_{LLM}$; 
$\varphi(model_i)$: the energy efficiency metric of $model_i$ in comparison with the base model;
$\rho(model_i)$: the performance metric of $model_i$ and defined as $\rho(model_i) = \frac{1}{N}\sum_{k=1}^N Metric_k$; $\forall Metric \in model_i$; $N$ is the total number of the metric used to evaluate model performance;
and $w$: a weighting factor between $\rho$ and $\varphi$ that determines the relative importance of performance and energy efficiency.

In the inference phase, the top $k$ LLMs based on the ranking criteria described above will be used to perform the task defined for the dataset.
To reduce energy consumption, we also explore the pruning technique with different pruning ratios (e.g., 10\%, 30\%) and pruning patterns on both magnitude-based and structured methods.
By systematically varying the pruning ratio and pruning pattern, we can gather comprehensive metrics which help to reduce the size of the model and computational complexity to achieve the desired trade-off between energy efficiency and accuracy.

The evaluation of pruned models will be conducted using the same performance metrics employed for the baseline model, non-pruned model.
These include well-known metrics for evaluating the quality of LLM outputs, including BERT \cite{zhang2019bertscore}, METEOR \cite{banerjee2005meteor}, BLEU \cite{papineni2002bleu}, ROUGE \cite{lin2004rouge} and cosine similarity.
This helps us assess the impact of pruning on both energy consumption and model quality.
By analyzing the energy efficiency and performance metrics of the pruned models compared to the baseline model, we can determine the balance pruning ratio or pruning pattern for achieving both energy efficiency and model quality.
Therefore, the analysis provides valuable insights into how pruning can be leveraged to enhance the energy efficiency for LLM without significantly compromising task performance.
At the end of the pipeline, a ranked collection of LLMs and their corresponding metrics will be assembled to assist report generation and provide valuable insights for selecting the right compromise settings between energy efficiency and model performance for the target application.

Our pipeline is evaluated on a high-performance computing server located at GAIA-Ericsson in Montreal.
The hardware environment consisted of 4 NVIDIA A30 GPUs, which were responsible for the bulk of the computations during training and inference, especially in energy-intensive tasks such as model pruning and quantization.
The CPU was an Intel(R) Xeon(R) Silver 4309Y with 32 cores running at 2.80GHz-sharedly used: for loading the data and some data preprocessing.
Besides this, it had been fitted with 256GB of RAM, thus giving quite sufficient memory when working with large-sized data and model parameters during training.
Experiments were conducted on Ubuntu OS; deep learning workflows were powered with standard frameworks like PyTorch, optimized for GPU acceleration.

\section{Results and Findings}

This section outlines our experiments with state-of-the-art LLMs, LLAMA-3 \cite{dubey2024llama} from Meta and Gemma \cite{team2024gemma} from Google, utilizing the proposed pipeline described in the previous section. 
With their substantial architectural advancements and performance enhancements, LLAMA-3 (8 billion parameters) and Gemma (7 billion parameters) serve as excellent candidates for testing strategies such as quantization and pruning.
However, it is important to note that the pipeline is designed to be versatile and can be applied to any LLM.

\begin{figure} [ht]
    \centering

    \begin{subfigure}[t]{1\columnwidth}
        \centering
    \includegraphics[width=0.85\columnwidth]{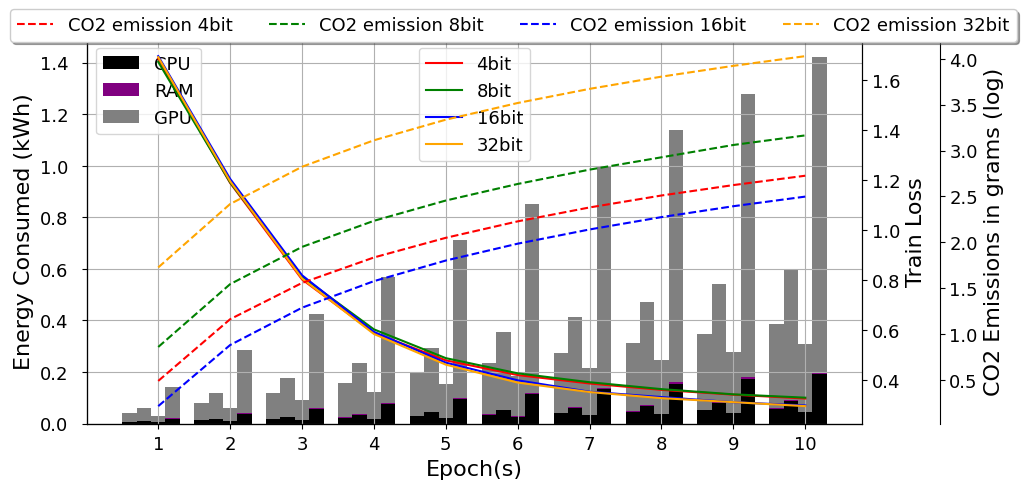}
    \vskip-5pt
    \caption{LLAMA3 - RCA dataset.}
    \end{subfigure}
    ~
    \begin{subfigure}[t]{1\columnwidth}
        \centering
    \includegraphics[width=0.85\columnwidth]{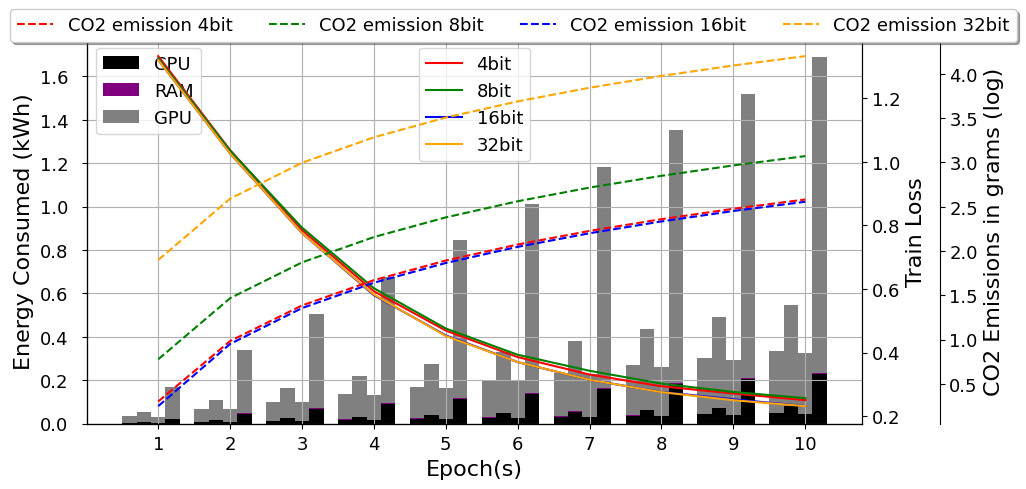}
    \vskip-5pt
    \caption{LLAMA3 - SAF dataset.}
    \end{subfigure}

    ~
    \begin{subfigure}[t]{1\columnwidth}
        \centering
    \includegraphics[width=0.85\columnwidth]{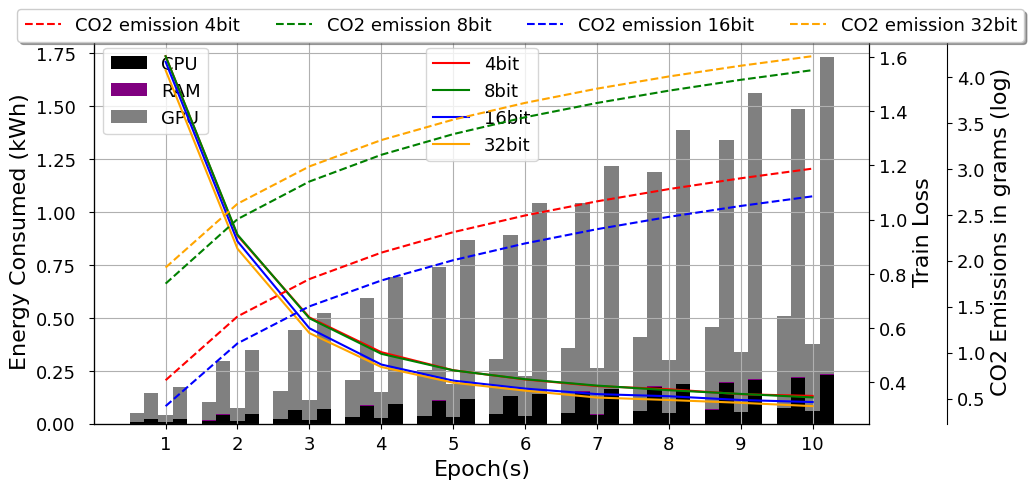}
    \vskip-5pt
    \caption{GEMMA - RCA dataset.}
    \end{subfigure}
    
    ~
    \begin{subfigure}[t]{1\columnwidth}
        \centering
    \includegraphics[width=0.85\columnwidth]{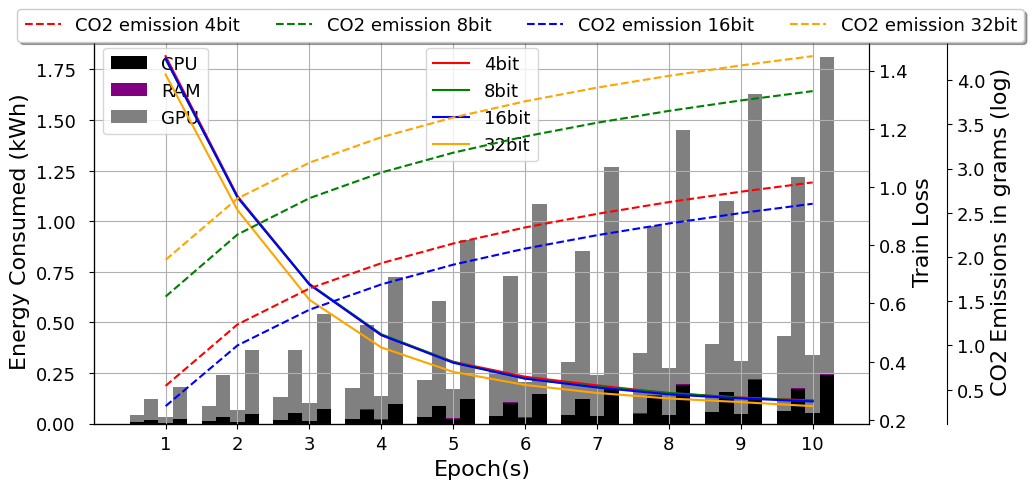}
    \vskip-5pt
    \caption{GEMMA - SAF dataset.}
    \end{subfigure}
    
    \caption{Energy consumption vs. training loss on each epoch\protect\footnotemark[1].}
    \label{fig:fine-tune-rca}
    \vskip-5pt
\end{figure}

As described in the previous section, this experiment explores quantization techniques at the fine-tuning step to reduce the computational and memory demands of LLM, therefore implying achieving energy efficiency.
Since our objective is to investigate energy efficiency in LLMs for RCA tasks, we set the weight factor $w = 0.7$, prioritizing energy efficiency gains over model performance in our evaluation.
Fig.\ref{fig:fine-tune-rca} illustrates the results of our experiment in the fine-tuning step, which evaluates the impact of different quantization levels, ranging from $2^2$-bit (4-bit) to $2^5$-bit (32-bit).
This technique offers a potential model size reduction of up to 47.98\% at other lower-bit models when compared to a 32-bit model.
However, as demonstrated by our results in a downstream application, the energy efficiency and performance trade-offs are more nuanced than initially theorized.
Across the quantization levels, 16-bit precision proved to be the most energy-efficient at all epochs during the fine-tuning process, followed closely by 4-bit, with 32-bit and 8-bit showing significantly higher energy consumption on both datasets used in our experiments.
In there, most of the energy consumption occurred in the GPU, which consistently uses around 6 times more energy compared to the CPU during the process.
Compared to the GPU, RAM usage has a minor impact on overall energy consumption, highlighting GPU is the dominant factor in computational energy costs.

\vskip-13pt

\footnotetext[1]{For each bar group, from left to right are 4-bit, 8-bit, 16-bit, and 32-bit.}

Fine-tuning performance varied depending on the quantization level at which models were performing during fine-tuning.
In the detailed training loss, the 16-bit model and the 32-bit model were almost identical, with only negligible changes throughout the epochs.
That means 16-bit precision performance is close to 32-bit performance, with energy efficiency; the best case reaches up to 80\% in energy efficiency.
While the performances of the 8-bit and 4-bit models are also similar to each other during the same range of epochs, they tend to show a bit higher training loss compared to 16-bit and 32-bit models.
Also, the performance difference did exist between the 16-bit/32-bit models and the 8-bit/4-bit models, though small, and the 32-bit model was better than its low-precision version.
Therefore, it underlines that the 16-bit model gives the best trade-off considering the power consumption and performance for the current task and hardware in this phase.

\begin{figure}[ht]
    \centering

     \begin{subfigure}[t]{1\columnwidth}
        \centering
    \includegraphics[width=0.85\columnwidth]{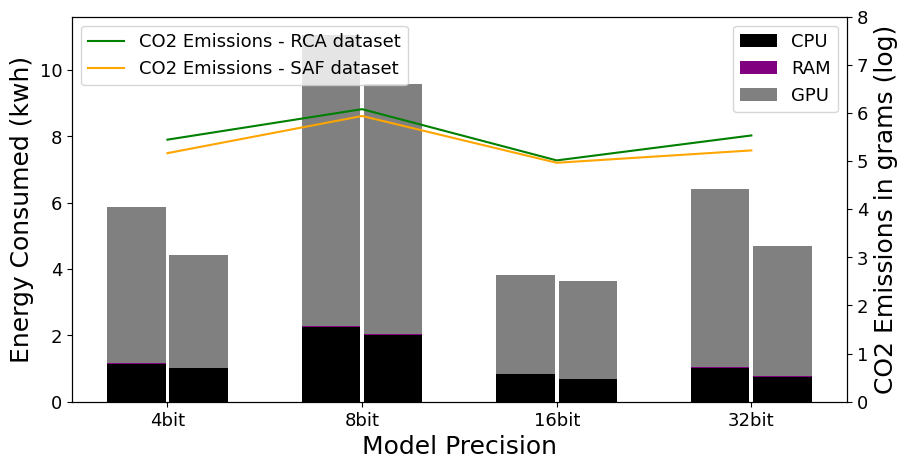}
    \caption{LLAMA3.}
    \end{subfigure}
   ~

    \begin{subfigure}[t]{1\columnwidth}
        \centering
    \includegraphics[width=0.85\columnwidth]{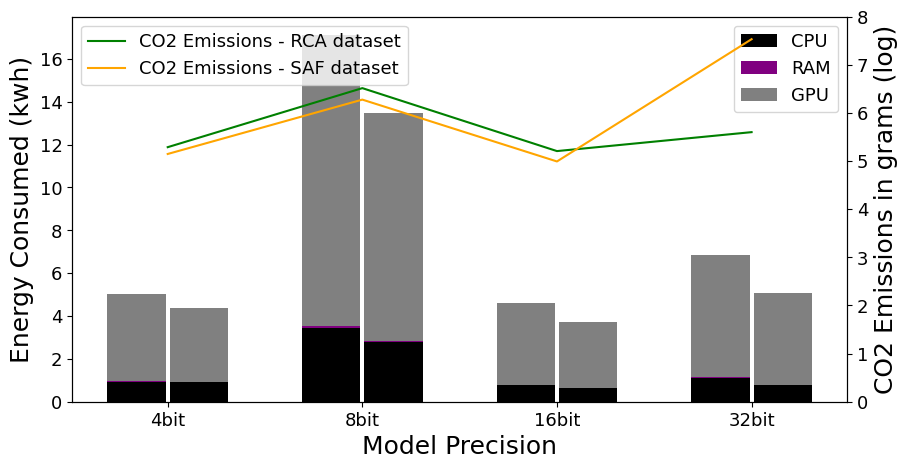}
    \caption{GEMMA.}
    \end{subfigure}
    
    \caption{Energy consumption vs. carbon emission vs. model precision in the inference phase\protect\footnotemark[2].}
    \label{fig:inference_energy_consumption}

\end{figure}

\footnotetext[2]{For each bar group, from left to right are RCA and SAF datasets.}

\begin{figure}[ht]
    \centering

    \begin{subfigure}[t]
        {1\columnwidth}
        \centering
        \includegraphics[width=0.85\columnwidth]{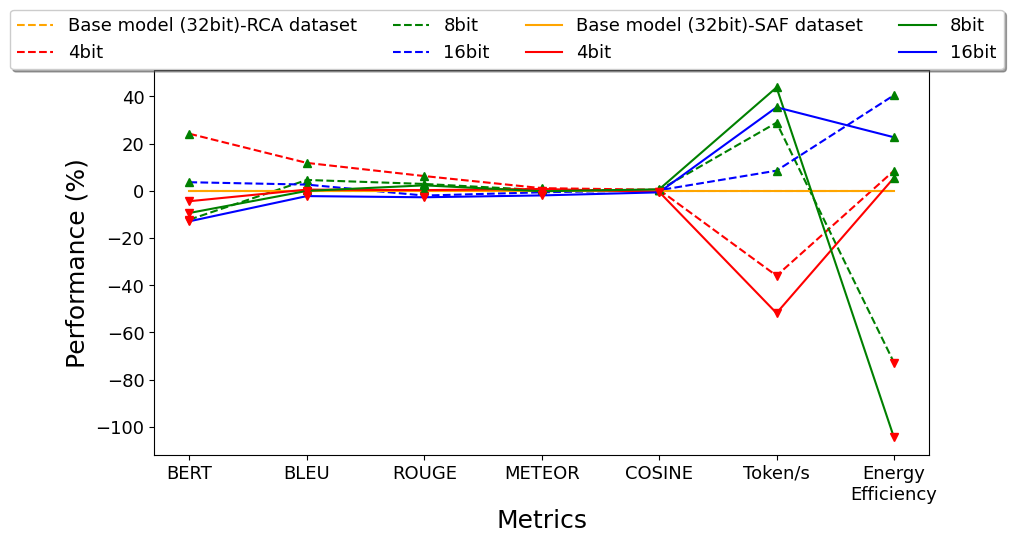}
        \vskip-5pt
        \caption{LLAMA3.}
    \end{subfigure}

    ~

    \begin{subfigure}[t]
        {1\columnwidth}
        \centering
        \includegraphics[width=0.85\columnwidth]{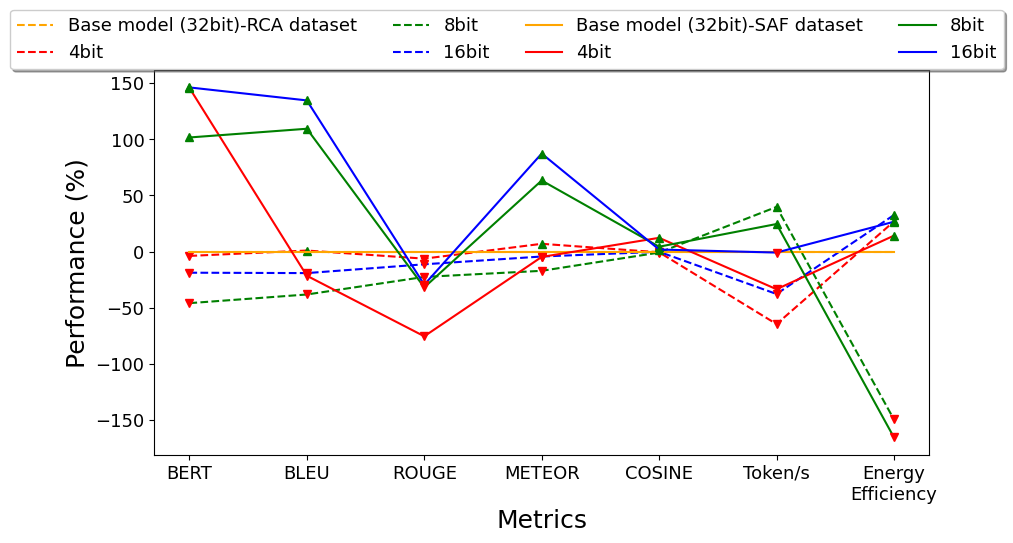}
        \vskip-5pt
        \caption{GEMMA.}
    \end{subfigure}

    \caption{Impact of quantization levels\protect\footnotemark[3].}
    \label{fig:model_llama_quantization_impact}
    \vskip-12pt
\end{figure}
\footnotetext[3]{Green triangle indicates better results in comparison to the base model.}

\begin{figure}[ht]
    \centering
    \includegraphics[width=1\columnwidth]{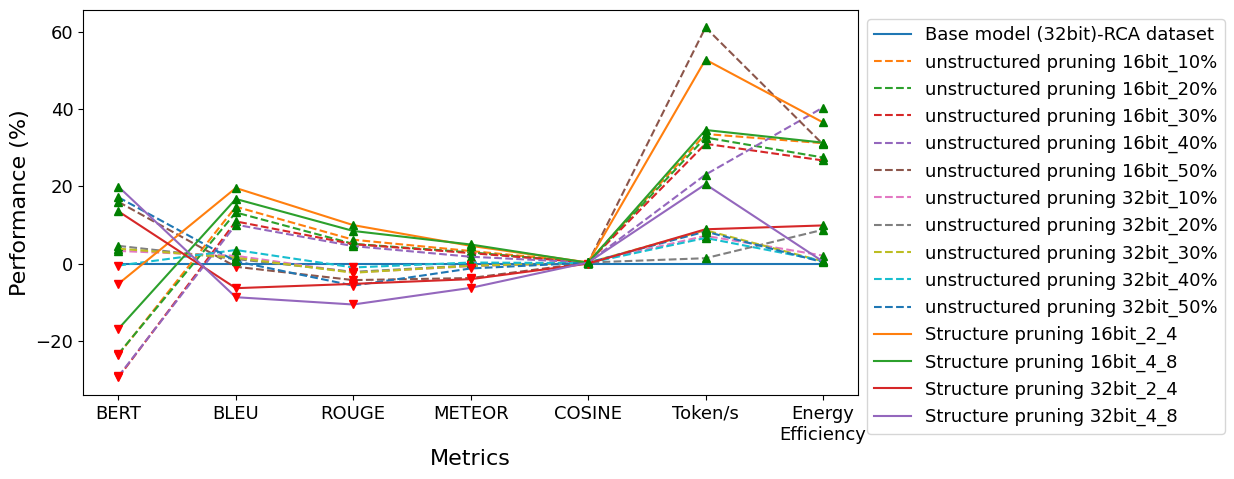}
    \caption{LLAMA3: impact of unstructured-base pruning and structured pruning\protect\footnotemark[3].}
    \label{fig:llama-pruning-rca}
    \vskip-15pt
\end{figure}    

Turning to the results from the inference phase, Fig.~\ref{fig:inference_energy_consumption} and Fig.~\ref{fig:model_llama_quantization_impact} highlight the effects of quantization levels during the inference phase.
Similar to the fine-tuning phase, the 16-bit model stands out as one of the top candidates for energy efficiency (reduction up to 40.5\% overall), offering a good compromise with model performance.
In detail, the BERT score drops slightly while other metrics show a slight improvement.
The 8-bit model demonstrates an improvement across most metrics, but this comes with a substantial rise in both energy consumption and the number of tokens generated per second.
Thus, this suggests that it can maintain accuracy while using significantly fewer bits in a downstream application.
However, this comes at a significant cost to energy efficiency, resulting in higher energy consumption and slower token generation.
The comparison underscores that while the 8-bit model offers improved performance over the 32-bit model in the same task, the trade-off in resource usage needs to be carefully evaluated, especially when aiming for energy efficiency.
One of the reasons for this is quantization also introduces quantization errors into the model, which adds extra computational cost as the model works to recover the weights during inference \cite{zhao2024atom}.
Another contributing factor is the lack of hardware support for lower-bit quantization on many GPUs, which are typically optimized for 16-bit floating-point operations and cause hardware-unfriendly memory access \cite{xia2024quant}.
This mismatch means that while the model can run at 8-bit precision, the hardware is not fully optimized for it, leading to inefficiencies.
As a result, the GPU must expend extra resources to compensate, further increasing energy consumption during inference.

Moving to the pruning techniques applied after fine-tuned LLM, the results in Fig.\ref{fig:llama-pruning-rca} demonstrate that both unstructured and structured pruning methods effectively reduce energy consumption.
In fact, by eliminating redundant weights or unimportant weights, these techniques simplify the model, making it more efficient without significantly reducing performance in a downstream task.
As shown in Fig. \ref{fig:llama-pruning-rca}, applying unstructured pruning to the original 32-bit model yields significant energy efficiency improvements, achieving up to 38\% gains across all pruning ratios (10\% to 50\%). Notably, this enhancement comes at the cost of only around 8.6\% reduction in performance (on average across metrics) in the worst-case scenario.
In addition, when utilizing the more efficient 16-bit precision, the model achieves energy savings of up to 40.42\% compared to the 32-bit version, with only around a 3\% performance drop on average across metrics compared to the unpruned 16-bit model, while achieving approximately 8\% better performance compared to the unpruned 32-bit version.
Similar to unstructured pruning, we evaluated structured pruning techniques by applying them to both the 32-bit and 16-bit models, utilizing 2:4 and 4:8 sparsity patterns.
Overall, structured pruning applied to these models consistently delivered energy efficiency improvements across all configurations. 
However, the 32-bit model showed more noticeable performance degradation on both 2:4 and 4:8 sparsity patterns compared to unstructured pruning and the baseline 32-bit model.
In contrast, the 16-bit model with structured pruning significantly enhanced performance across metrics such as BLEU, ROUGE, METEOR, cosine similarity, and the number of tokens generated per second when compared to the unpruned 16-bit model and the unstructured-pruned models.
Impressed is the 2:4 sparsity pattern on the 16-bit model which increases both model performance and energy efficiency in comparison to the baseline model and 16-bit model without pruning.
This suggests that applying a combination of 16-bit quantization and 2:4 structured pruning achieves the best compromise between model performance and energy efficiency, making it an ideal approach for toward sustainable AI system on our hardware configuration.

\section{Conclusion}

In this research, we investigated the trade-off between energy efficiency and performance for LLM in a downstream application deployed in communication networks.
We proposed a novel pipeline that integrates quantization and pruning techniques to assist in addressing compromise between both above factors. 
In addition, our pipeline was evaluated using two real-world datasets focusing on RCA application and question-answer tasks within communication networks.
Our findings demonstrate,
while quantization techniques can significantly reduce the model size and memory footprint, the optimal level of quantization for achieving the best balance depends on the specific task and hardware configuration.
On the other hand, pruning techniques also offer some potential for energy efficiency gains.
However, aggressive pruning can lead to a significant degradation in model performance as demonstrated in our results.
The right combination of quantization and pruning techniques can achieve energy efficiency while also enhancing model performance.

In future work, we aim to extend the evaluation of our pipeline to different domains, such as economics, and apply it to various tasks, including summarization.
Additionally, we plan to assess the performance of our pipeline on low-power devices and heterogeneous hardware platforms, such as FPGAs and TPUs.

\section*{Acknowledgement}

We are grateful to Adel Larabi, Jungyeon Baek and Karthikeyan Premkumar at Ericsson-GAIA, Montréal for their support throughout this project.

\bibliographystyle{unsrt2authabbrvpp}
\bibliography{icc_2025}

\end{document}